\documentclass[11pt]{article}

\usepackage{indentfirst}
\usepackage{graphicx}
\usepackage{longtable}
\usepackage{supertabular}
\usepackage{amsmath}
\usepackage{amssymb}
\usepackage{color}
\usepackage{amstext}
\usepackage{cases}
\usepackage{float}
\usepackage{threeparttable}
\usepackage{booktabs}
\usepackage[margin=.8in]{geometry}
\usepackage{setspace}
\usepackage{multirow}
\usepackage{array}
\usepackage{mathrsfs}

\begin{document}

\title{{\Large\bf {Conditions for none to be whipped by `Rank and Yank' under the majority rule}}}

\author{Fujun Hou \thanks{Email: houfj@bit.edu.cn.} \\
School of Management and Economics\\
Beijing Institute of Technology\\
Beijing, China, 100081
}
\date{\today}
\maketitle

\begin{abstract}
`Rank and Yank' is practiced in many organizations. This paper is concerned with the condtions for none to be whipped by `Rank and Yank' when the evaluation data under each criterion are assumed to be ordinal rankings and the majority rule is used. Two sufficient conditions are set forth of which the first one formulates the alternatives indifference definition in terms of the election matrix, while the second one specifies a certain balance in the probabilities of alternatives being ranked at positions. In a sense, `none to be whipped' means that the organization is of stability. Thus the second sufficient condition indicates an intrinsic relation of balance and organization stability. In addition, directions for future research are put forward.

\end{abstract}

\setlength{\unitlength}{1mm}

\section{Introduction}

We consider on what conditions the `Rank and Yank' will not whip anyone in an organization provided that the evaluation data under each criterion are ordinal rankings and the simple majority rule is used as the aggregation rule.

`Rank and Yank', also called `Up or Out policy', `Forced Ranking', or `Vitality Curve' popularized by General Electric and later adopted by many organizations, refers to a forced distribution rating system (FDRS) that rewards the top employees and removes the bttom employees (in certain ratios, for instance, 20:70:10) each period (in a certain frequency, for instance, every year or every 4 year). This policy or system has been extensively investigated by the academia. Topics cover its advanges and disadvantages from various prospectives (see, e.g., Dominick 2009, Giumetti et al. 2015, Stewart et al. 2010, and many others). 

In a `Rank and Yank' scenario, the performances of the employees of a given organization are aggregated into a social preference that reflects the organization's general opinion based on which the employees are ranked. It is obvious when the employees under stake are rated with respect to a single criterion and the employees perform same good, then nobody will be whipped by the `Rank and Yank' policy. This is a simple but trivial case. When the employees are rated with respect to multiple criteria, and the evaluation data under each criterion are assumed of ordinal rankings, the situations can be a bit more complicated in the sense that an employee may perform better with respect to one criterion while worse with respect to another. This paper does not touch on such topics as the policy's ethical issues, reasonability or efficiency. Rather, we concentrate on a certain situation of the policy, that is, even though the policy works well but none will be rewarded or punished given that the evaluation data under each criterion are ordinal rankings and the simple majority rule is used. When this phenomenon (i.e., `none to be whipped') is to be observed, our question is, what the conditions will be?

This question is different from two much concerned topics in social choice theory, one of which is known as `social preference transitivity under the majority rule', and the other is `social choice set existence under the majority rule' (for details see, e.g., Arrow 1951, Sen 1970, and Gaertner 2009). This is evidenced by the following illustration where 3 alternatives $\{x_1,x_2,x_3\}$ are evaluated (the terminology $x\sim y$ stands for `$x$ is indifferent to $y$', and $x\succ y$ stands for `$x$ is more preferred to $y$'):
\begin{itemize}
\item On the one hand, if the final aggregation outcome is $x_1\sim x_2\sim x_3$, then the social preference is transitive and the social choice set is not empty. In this case will nobody be whipped by Rank and Yank.
\item One the other hand, if the final aggregation outcome is $x_1\succ x_2\succ x_3\succ x_1$, then the social preference is not transitive and the social choice set does not exist either. However, `none to be whipped' still happens in this particular case.
\end{itemize}
Therefore, the occurence of `none to be whipped' is independent of the transitivity of the social preference or the existence of the social choice set.

We will present two sufficient conditions. To this end, the paper is organized as follows. Section 2 includes our assumptions and notations. Some aggregation outcomes possible for `none to be whipped' are listed in Section 3. Section 4 presents a sufficient condition in terms of the election matrix whose entries represent the numbers of criteria under which one alternative being preferred to another. In Section 5, we first introduce a matrix called preference probability map (PPM), whose entries stand for the probabilities of alternatives occupying ranking positions. We then obtain a sufficient condition in terms of the sum and mean matrix of the PPMs. Some examples are also examined. And Section 6 includes our concluding remark and put forward some possible directions for future research.

\section{Assumption and notation}

We henceforth call the employees as alternatives. The discussion rests on the following \textit{assumptions}.

\begin{itemize}
\item []\textbf{Assumption I} The alternatives are finite, evaluated under multiple criteria and the criteria are dealt with equally; 
\item []\textbf{Assumption II} The evaluation data under each criterion are collected as ordinal rankings;
\item []\textbf{Assumption III} The simple majority rule is used as the aggregation rule;
\item []\textbf{Assumption IV} When to rank alternatives with respect to one criterion, those in a tie will occupy common consecutive positions with an equal probability;
\item []\textbf{Assumption V} The evaluation data under different criteria are independent in the sense that, even though an alternative performs good regarding one criterion, he/she may not necessarily perform good regarding another criterion.
\end{itemize}

The \textit{significance} of the above assumptions lie in that:
\begin{itemize}
\item [$\diamond$] The first three assumptions imply that the 'Rank and Yank' problem under consideration can be treated as a social choice problem where {\em the ordinal ranking of the alternatives under a criterion is viewed as provided by a voter};
\item [$\diamond$] The fourth assumption indicates a probability mapping of each ordinal ranking, which will be described in Section 5; 
\item [$\diamond$] The last assumption ensures the feasibility of probability addition in the discussion.
\end{itemize}

To conduct the discussion, the following \textit{notations} will be used throughout the paper.
\begin{itemize}
\item [-] $x_i$ denote alternatives (employees), where $i=1,2,\ldots, m$ and $1<m<+\infty$. Accordingly, $X=\{x_1,x_2,\ldots,x_m\}$ stands for the alternative set.
\item [-] $c_j$ denote criteria (voters), where $j=1,2,\ldots, n$ and $1<n<+\infty$. Accordingly, $\{c_1,c_2,\ldots,c_n\}$ stands for the criterion (voter) set.
\item [-] Possible evaluation data of alternatives $x_i$ and $x_k$ with respect to criterion $j$ can be one of $x_i\succ_j x_k$, $x_i\sim_j x_k$ and $x_i\prec_j x_k$, which stand for ``$x_i$ is more preferred to $x_k$", ``$x_i$ is indifferent to $x_k$" and ``$x_i$ is less preferred to $x_k$", respectively. When we speak of social preferences over alternatives $x_i$ and $x_k$, we just use similar notations but without any subscript attached to `$\succ$', `$\sim$' or `$\prec$'.  
\item [-] $N(x_i\succ_j x_k)$ and $N(x_i\sim_j x_k)$ denote the numbers of criteria satisfying $x_i\succ_j x_k$ and $x_i\sim_j x_k$, respectively.
\item [-] The majority rule means that $x_i\succ x_k$ iff $N(x_i\succ_j x_k)>N(x_k\succ_j x_i)$, and $x_i\sim x_k$ iff $N(x_i\succ_j x_k)=N(x_k\succ_j x_i)$.
\end{itemize}

We note that the majority rule in this paper belongs to a `simple majority rule' rather than an `absolute majority rule'. For knowledge of the difference, readers may refer to Arrow (1951), Sen (1970), and Gaertner (2009).

\section{Aggregation outcomes leading to `none to be whipped'}

A fundamental problem in our topic is to identify what final aggregation outcomes will lead to the occurrence of `none to be whipped'. When the majority rule does not deliver a distinguishable result, indifference or cyclicity may hold over the whole alternative set. We list some outcomes of this sort as follows.

\begin{itemize}
\item [(1)] The majority rule yields a final aggregation outcome that all alternatives are indifferent to each other. For instance, when to evaluate 4 alternatives, the particular case of this kind refers to the case where the final collective outcome is $x_1\sim x_2\sim x_3\sim x_4$.
\item [(2)] The majority rule yields a final aggregation outcome that cyclicity holds over the whole alternative set. For instance, when to evaluate 5 alternatives, the particular case of this kind can be the case where the final collective outcome is $x_1\succ x_2\succ x_3\succ x_4\succ x_5\succ x_1$.
\item [(3)] The majority rule yields a final aggregation outcome that is similar to an indifference line including some cycles. For instance, the final collective outcome is such like $x_1\sim\{x_2\succ x_3\succ x_4\}$.
\item [(4)] When the majority rule yields a final aggregation outcome that a global cycle occurs with some local indifference and/or cyclicity mixed in the cycle, 'none to be whipped' will occur. For illustration, the cases can be such as
$$\begin{array}{ll}
 x_1\succ \{x_2\sim x_3\sim x_4\}\succ x_5\succ x_1,\\
 x_1\succ \{x_2\succ x_3\succ x_4\succ x_2\}\succ x_5\succ x_1,\\
 x_1\succ \{x_2\sim x_3\sim x_4\}\succ x_5\succ \{x_6\succ x_7\succ x_8\succ x_9\succ x_6\}\succ x_{10}\succ x_1, \mbox{etc}.
\end{array}
$$
\end{itemize}

In this paper, we will make an attempt to formulate mathematically some sufficient conditions relevant to the first three cases.

\section{A sufficient condition in terms of election matrix}

Binary comparison matrices are widely used in social choice literature whose elements can be assumed of various meannings. For our analytical purpose in this section, we use the one called \textit{election matrix} $M=(a_{ik})_{m\times m}$ whose entry $a_{ik}$ represents the number of times alternative $x_i$ is ranked before aternative $x_k$ across the $n$ criteria (Levenglick 1975). With this interpretation of the entries of the election matrix, the following condition becomes apparent.

\textbf{Sufficient condition in terms of election matrix} If the election matrix $M=(a_{ik})_{m\times m}$ is symmetric, i.e., $a_{ik}=a_{ki}$, then none will be whipped by the `Rank and Yank' under the majority rule.

\textbf{Proof} Since $a_{ik}=N(x_i\succ_j x_k)$ hence $a_{ik}=a_{ki}$ means $N(x_i\succ_j x_k)=N(x_k\succ_j x_i)$, we thus have $x_i\sim x_k$, $\forall i,k$, under the majority rule. Therefore, in this case all alternatives are indifferent to each other and none will be whipped by the `Rank and Yank'.

\section{A sufficient condition in terms of preference probability map}

We start with two definitions which will prove to be useful for the representation of ordinal rankings. Then we present a sufficient condition in terms of one of them.

\textbf{Preference map} (Hou 2015a, 2015b; Hou \& Triantaphyllou 2019) The \emph{preference map} is a sort of preference sequence corresponding to individual $j$'s weak ordering over the alternative set $X=\{x_1,x_2,\ldots,x_m\}$, denoted $PM^{(j)}=[PM_i^{(j)}]_{m\times 1}$, such that
$$PM_i^{(j)}=\{|A_i^{(j)}|+1,|A_i^{(j)}|+2,\ldots,|A_i^{(j)}|+|B_i^{(j)}|\},\eqno(1)$$
where $|\centerdot|$ represents the cardinality of a set;
$A_i^{(j)}$ represents the \textit{predominance set} of alternative $x_i$ according to individual $j$'s preference, i.e., $A_i^{(j)}=\{x_q\mid x_q\in X, x_q \succ_j x_i\}$; and $B_i^{(j)}$ represents the \textit{indifference set} of alternative $x_i$, i.e., $B_i^{(j)}=\{x_q\mid x_q\in X,x_q \sim_j x_i\}$.

For example, we consider the orderings of $x_1\succ x_2\succ x_3\succ x_4$, $x_1\sim x_2\succ x_3\sim x_4$, $x_1\succ x_2\sim x_3\sim x_4$ and $x_1\sim x_2\sim x_3\sim x_4$. Their preferences maps are
$$
\begin{array}{c}
{\begin{array}{cc}
{\begin{array}{c} x_1\\
x_2\\
x_3\\
x_4
\end{array}}&
{ \left[\begin{array}{c}
\{1\}\\
\{2\}\\
\{3\}\\
\{4\}\end{array}\right],
\left[\begin{array}{c}\{1,2\}\\
\{1,2\}\\
\{3,4\}\\
\{3,4\}\end{array}\right],
\left[\begin{array}{c}\{1\}\\
\{2,3,4\}\\
\{2,3,4\}\\
\{2,3,4\}\end{array}\right]\mbox{~and~}
\left[\begin{array}{c}\{1,2,3,4\}\\
\{1,2,3,4\}\\
\{1,2,3,4\}\\
\{1,2,3,4\}\end{array}\right],
}
\end{array}}
\end{array}
$$
respectively.

From the definition of the preference map one can see that, a preference map is an `almost vector' whose entries are sets containing the alternatives' possible ranking position or positions, that is, each alternative is mapped to its ranking position or positions. If two alternatives are in a tie, then their corresponding entries include same numbers which indicate that those alternatives in a tie will occupy same ranking positions. This characteristic of the preference map reflects the requirements in Assumption IV that those alternatives ``in a tie will occupy common consecutive positions". In order to reflect exactly the ``probability" information required in Assumption IV, however, we need another concept as described below.

\textbf{Preference possibility map (Hou 2022)} Let $PM=[PM_i]_{m\times 1}$ be a preference map of a weak ordering. The preference possibility map (for short, PPM) corresponding to that weak ordering is defined by a $m\times m$ matrix $[PPM_{i,k}]_{m\times m}$ such that
$$
{PPM_{i,k}}=\begin{cases}
\frac{1}{|PM_i|}, & \text{if $k\in PM_i$},\\
0, & \text{otherwise}.
\end{cases}
\eqno(2)$$

One can see from Eq.(2) that the PPM reflects exactly the requirement in Assumption IV that those alternatives ``in a tie will occupy common consecutive positions with an equal probability". We note that this sort of strategy (i.e., alternatives in a tie will occupy common consecutive positions with an equal probability) can be traced back at least as early as Black (1976) when to assign Borda scores to those alternatives in a tie.

In our previous example, the corresponding PPMs would be
$$
\begin{array}{c}
\begin{array}{c}{\hspace{0.7cm}\begin{array}{ccccc}1\hspace{0.3cm}2\hspace{0.4cm}3\hspace{0.4cm}4
&\hspace{1.0cm}1\hspace{0.7cm}2\hspace{0.7cm}3\hspace{0.7cm}4 &\hspace{1.0cm}1\hspace{0.7cm}2\hspace{0.7cm}3\hspace{0.7cm}4 &\hspace{1.8cm}1\hspace{0.7cm}2\hspace{0.7cm}3\hspace{0.7cm}4
\end{array}}\\
{\begin{array}{cc}
{\begin{array}{c} x_1\\
x_2\\
x_3\\
x_4
\end{array}}&
{ \left[\begin{array}{cccc}
1 &  0 &  0 &  0 \\
0 &  1 &  0 &  0 \\
0 &  0 &  1 &  0 \\
0 &  0 &   0&  1
\end{array}\right],
\left[\begin{array}{cccc}
1/2 &  1/2 &  0 &  0 \\
1/2 &  1/2 &  0 &  0  \\
0 &  0 &  1/2  &  1/2 \\
0 &  0 &  1/2  &  1/2 
\end{array}\right],
\left[\begin{array}{cccc}
1 &  0 &  0 &  0 \\
0 &  1/3 &  1/3 &  1/3 \\
0 &  1/3 &  1/3 &  1/3 \\
0 &  1/3 &  1/3 &  1/3
\end{array}\right]\mbox{~and~}
\left[\begin{array}{cccc}
1/4 &  1/4 &  1/4 &  1/4 \\
1/4 &  1/4 &  1/4 &  1/4 \\
1/4 &  1/4 &  1/4 &  1/4 \\
1/4 &  1/4 &  1/4 &  1/4
\end{array}\right],
}
\end{array}}
\end{array}
\end{array}
$$
respectively.

From its definition we know that a PPM has the following properties.

\textbf{Property of PPM} Suppose $[PPM_{i,k}]_{m\times m}$ is a PPM. We have
\begin{itemize}
\item $0\leq PPM_{i,k}\leq 1, \forall i,k$;
\item $\sum_{i=1}^{m}PPM_{i,k}=1$;
\item $\sum_{k=1}^{m}PPM_{i,k}=1$.
\end{itemize}

The above property of PPM indicates that, an alternative occupies at least one ranking position, meanwhile a ranking position is occupied at least by one alternative. Actually, this is a matter of fact that a PPM represents an ordinal ranking.

The following definition is crucial for the analysis in this section.

\textbf{Definition 1} Suppose that $[PPM_{i,k}^{(j)}]_{m\times m}$ are PPMs, which correspond to the alternatives' ordinal rankings with respect to criteria $j$, $j=1,2,\ldots,n$. Their \textit{sum matrix} and \textit{mean matrix} are defined by
$$\Sigma PPM=[n_{i,k}]_{m\times m}=\left[\Sigma_{j=1}^{n}PPM_{i,k}^{(j)}\right]_{m\times m}$$
and $$\overline{\Sigma PPM}=[\overline{p}_{i,k}]_{m\times m}=\left[\frac{1}{n}\Sigma_{j=1}^{n}PPM_{i,k}^{(j)}\right]_{m\times m},$$
respectively.

Since $PPM_{i,k}^{(j)}$ is interpreted as the probability of alternative $x_i$ ranking at position $k$ under criterion $j$, and the probabilities across the criteria are assumed independent of each other (Assumption V), thus $\overline{p}_{i,k}$ represents the average probability of the society's ranking alternative $x_i$ at position $k$. From the property of PPM, the following properties hold respectively for mean matrix and sum matrix.

\textbf{Property of mean matrix} Suppose $[\overline{p}_{i,k}]_{m\times m}$ is a mean matrix. We have
\begin{itemize}
\item $0\leq\overline{p}_{i,k}\leq 1, \forall i,k$;
\item $\sum_{i=1}^{m}\overline{p}_{i,k}=1$;
\item $\sum_{k=1}^{m}\overline{p}_{i,k}=1$.
\end{itemize}

\textbf{Property of sum matrix} Suppose $[n_{i,k}]_{m\times m}$ is a sum matrix. We have
\begin{itemize}
\item $n_{i,k}\geq 0, \forall i,k$;
\item $\sum_{i=1}^{m}n_{i,k}=n$;
\item $\sum_{k=1}^{m}n_{i,k}=n$.
\end{itemize}

\textbf{Lemma 1} The number of criteria under which alternative $x_i$ is ranked at position $k$ is 
$$n_{i,k}=\Sigma_{j=1}^{n}PPM_{i,k}^{(j)}=n\times \overline{p}_{i,k}.$$

\textbf{Proof} According to classical definition of probability, we know $\overline{p}_{i,k}=\frac{n_{i,k}}{n}$. Given $\overline{p}_{i,k}=\frac{1}{n}\Sigma_{j=1}^{n}PPM_{i,k}^{(j)}$, we have $n_{i,k}=\Sigma_{j=1}^{n}PPM_{i,k}^{(j)}=n\times \overline{p}_{i,k}$. Q.E.D

The following result can be established.

\textbf{Lemma 2} If $\forall i,k\left(\overline{p}_{i,k}=\frac{1}{m}\right)$ holds for the mean matrix, then, none of the alternatives $\{x_1,x_2,\ldots,x_m\}$ will be whipped by the `Rank and Yank'.

\textbf{Proof} Since $\overline{p}_{i,k}$ stands for the social probability of alternative $x_i$ ranking at position $k$, thus when $\forall i,k\left(\overline{p}_{i,k}=\frac{1}{m}\right)$ holds for the mean matrix, any alternative would occupy any position with an identical probability. In this case, we know that either all the alternatives are indifferent to each other, or cyclicity holds over the whole alternative set. Given the condition, therefore, none will be whipped by the `Rank and Yank'. Q.E.D

Can Lemma 2 be generalized even further? We have the following result.

\textbf{Theorem 1 (A sufficient condition)} If the entries of the mean matrix satisfy a dual relation, that is,
$$\forall i(\overline{p}_{i,1}=\overline{p}_{i,n}, \overline{p}_{i,2}=\overline{p}_{i,n-1}, \ldots),\eqno(3)$$
or equivalently, the entries of the sum matrix satisfy the dual relation
$$\forall i(n_{i,1}=n_{i,n}, n_{i,2}=n_{i,n-1}, \ldots),\eqno(4)$$
then, none of the alternatives $\{x_1,x_2,\ldots,x_m\}$ will be whipped by the `Rank and Yank'.

\textbf{Proof} Given the condition of formula (3), we know that
\begin{itemize}
\item If $\forall i,k\left(\overline{p}_{i,k}=\frac{1}{m}\right)$, then from Lemma 2 we know that the theorem holds.
\item Otherwise, from formula (3) we know that the average ranking position of any alternative will be $\frac{m+1}{2}$, which indicates that the aggregation outcome will be an indifference line with or without some cycles in it. Thererefore, in this case none will be whipped. Q.E.D 
\end{itemize}

\textbf{Example 1} Consider $X=\{x_1,x_2,x_3,x_4\}$, and the evaluation data under 5 criteria are given as:
$$\begin{array}{ll}
 \text{Criterion 1}: & x_1 \succ_1 x_2 \sim_1 x_3 \sim_1 x_4,\\
 \text{Criterion 2}: & x_2 \sim_2 x_3 \sim_2 x_4 \succ_2 x_1,\\
 \text{Criterion 3}: & x_1 \sim_3 x_2 \succ_3 x_3 \sim_3 x_4,\\
 \text{Criterion 4}: & x_3 \sim_4 x_4 \succ_4 x_1 \sim_4 x_2,\\
 \text{Criterion 5}: & x_1 \sim_5 x_2 \sim_5 x_3 \sim_5 x_4.
\end{array}
$$
The corresponding PPMs are
$$
\begin{array}{c}
{\begin{array}{cc}
{\begin{array}{c} x_1\\
x_2\\
x_3\\
x_4
\end{array}}&
{ \left[\begin{array}{cccc}
1 &  0 &  0 &  0 \\
0 &  \frac{1}{3} &  \frac{1}{3} &  \frac{1}{3} \\
0 &  \frac{1}{3} &  \frac{1}{3} &  \frac{1}{3} \\
0 &  \frac{1}{3} &   \frac{1}{3} &  \frac{1}{3}
\end{array}\right],
\left[\begin{array}{cccc}
0 &  0 &  0 &  1 \\
\frac{1}{3} &  \frac{1}{3} &  \frac{1}{3} &  0  \\
\frac{1}{3} &  \frac{1}{3} &  \frac{1}{3}  &  0 \\
\frac{1}{3} &  \frac{1}{3} &  \frac{1}{3}  &  0 
\end{array}\right],
\left[\begin{array}{cccc}
\frac{1}{2} &  \frac{1}{2} &  0 &  0 \\
\frac{1}{2} &  \frac{1}{2} &  0 &  0 \\
0 &  0 &  \frac{1}{2} &  \frac{1}{2} \\
0 &  0 &  \frac{1}{2} &  \frac{1}{2}
\end{array}\right],
\left[\begin{array}{cccc}
0 &  0 &  \frac{1}{2} &  \frac{1}{2} \\
0 &  0 &  \frac{1}{2} &  \frac{1}{2} \\
\frac{1}{2} & \frac{1}{2} &  0 &  0 \\
\frac{1}{2} & \frac{1}{2} &  0 &  0
\end{array}\right],
\left[\begin{array}{cccc}
\frac{1}{4} &  \frac{1}{4} &  \frac{1}{4} &  \frac{1}{4} \\
\frac{1}{4} &  \frac{1}{4} &  \frac{1}{4} &  \frac{1}{4} \\
\frac{1}{4} &  \frac{1}{4} &  \frac{1}{4} &  \frac{1}{4} \\
\frac{1}{4} &  \frac{1}{4} &  \frac{1}{4} &  \frac{1}{4}
\end{array}\right].
}
\end{array}}
\end{array}
$$
The sum matrix is
$$
\begin{array}{c}
{\begin{array}{cc}
{\begin{array}{c} x_1\\
x_2\\
x_3\\
x_4
\end{array}}&
{ 
\left[\begin{array}{cccc}
\frac{7}{4} &  \frac{3}{4} &  \frac{3}{4} &  \frac{7}{4} \\
\frac{13}{12} &  \frac{17}{12} &  \frac{17}{12} &  \frac{13}{12} \\
\frac{13}{12} &  \frac{17}{12} &  \frac{17}{12} &  \frac{13}{12} \\
\frac{13}{12} &  \frac{17}{12} &  \frac{17}{12} &  \frac{13}{12}
\end{array}\right].
}
\end{array}}
\end{array}
$$
Condition (4) is fulfilled. Hence none will be whipped by `Rank and Yank' in this example. Actually, the final aggregation outcome is $x_1 \sim x_2 \sim x_3 \sim x_4$.

\textbf{Example 2} Consider $X=\{x_1,x_2,x_3,x_4\}$, and the evaluation data under 6 criteria are given as:
$$\begin{array}{ll}
 \text{Criterion 1}: & x_1 \succ_1 x_2 \succ_1 x_3 \succ_1 x_4,\\
 \text{Criterion 2}: & x_1 \succ_2 x_3 \succ_2 x_4 \succ_2 x_2,\\
 \text{Criterion 3}: & x_1 \succ_3 x_4 \succ_3 x_2 \succ_3 x_3,\\
 \text{Criterion 4}: & x_2 \succ_4 x_3 \succ_4 x_4 \succ_4 x_1,\\
 \text{Criterion 5}: & x_3 \succ_5 x_4 \succ_5 x_2 \succ_5 x_1,\\
 \text{Criterion 6}: & x_4 \succ_6 x_2 \succ_6 x_3 \succ_6 x_1.
\end{array}
$$
The corresponding PPMs are
\begin{footnotesize}
$$
\begin{array}{c}
{\begin{array}{cc}
{\begin{array}{c} x_1\\
x_2\\
x_3\\
x_4
\end{array}}&
{ \left[\begin{array}{cccc}
1 &  0 &  0 &  0 \\
0 &  1 &  0 &  0 \\
0 &  0 &  1 &  0 \\
0 &  0 &   0 &  1
\end{array}\right],
\left[\begin{array}{cccc}
1 &  0 &  0 &  0 \\
0 &  0 &  0 &  1 \\
0 &  1 &  0 &  0 \\
0 &  0 &   1 &  0 
\end{array}\right],
\left[\begin{array}{cccc}
1 &  0 &  0 &  0 \\
0 &  0 &  1 &  0 \\
0 &  0 &  0 &  1 \\
0 &  1 &   0 &  0
\end{array}\right],
\left[\begin{array}{cccc}
0 &  0 &  0 &  1 \\
1 &  0 &  0 &  0 \\
0 &  1 &  0 &  0 \\
0 &  0 &   1 &  0
\end{array}\right],
\left[\begin{array}{cccc}
0 &  0 &  0 &  1 \\
0 &  0 &  1 &  0 \\
1 &  0 &  0 &  0 \\
0 &  1 &   0 &  0 
\end{array}\right],
\left[\begin{array}{cccc}
0 &  0 &  0 &  1 \\
0 &  1 &  0 &  0 \\
0 &  0 &  1 &  0 \\
1 &  0 &   0 &  0
\end{array}\right].
}
\end{array}}
\end{array}
$$
\end{footnotesize}
The sum matrix is
$$
\begin{array}{c}
{\begin{array}{cc}
{\begin{array}{c} x_1\\
x_2\\
x_3\\
x_4
\end{array}}&
{ 
\left[\begin{array}{cccc}
3 &  0 &  0 &  3 \\
1 &  2 &  2 &  1 \\
1 &  2 &  2 &  1 \\
1 &  2 &  2 &  1 
\end{array}\right].
}
\end{array}}
\end{array}
$$
Condition (4) is fulfilled. Hence none will be whipped by `Rank and Yank' in this example. Actually, the final aggregation outcome is $x_1 \sim \{x_2 \succ x_3 \succ x_4\succ x_2\}$.

\textbf{Remark} The condition presented in Theorem 1 is merely a sufficient condition rather than a necessary one. We illustrate by an example whose data are taken from  Gaertner (2009, P.110). Consider $X=\{x_1,x_2,x_3\}$, and the evaluation data under 5 criteria are given as:
$$\begin{array}{ll}
 \text{Criterion 1}: & x_1 \succ_1 x_2 \succ_1 x_3,\\
 \text{Criterion 2}: & x_1 \succ_2 x_2 \succ_2 x_3,\\
 \text{Criterion 3}: & x_2 \succ_3 x_3 \succ_3 x_1,\\
 \text{Criterion 4}: & x_2 \succ_4 x_3 \succ_4 x_1,\\
 \text{Criterion 5}: & x_3 \succ_5 x_1 \succ_5 x_2.
\end{array}
$$
The corresponding PPMs are
$$
\begin{array}{c}
{\begin{array}{cc}
{\begin{array}{c} x_1\\
x_2\\
x_3
\end{array}}&
{ \left[\begin{array}{ccc}
1 &  0 &  0 \\
0 &  1 &  0 \\
0 &  0 &  1
\end{array}\right],
\left[\begin{array}{ccc}
1 &  0 &  0 \\
0 &  1 &  0 \\
0 &  0 &  1
\end{array}\right],
\left[\begin{array}{ccc}
0 &  0 &  1 \\
1 &  0 &  0 \\
0 &  1 &  0
\end{array}\right],
\left[\begin{array}{ccc}
0 &  0 &  1 \\
1 &  0 &  0 \\
0 &  1 &  0
\end{array}\right],
\left[\begin{array}{ccc}
0 &  1 &  0 \\
0 &  0 &  1 \\
1 &  0 &  0 
\end{array}\right].
}
\end{array}}
\end{array}
$$
The sum matrix is
$$
\begin{array}{c}
{\begin{array}{cc}
{\begin{array}{c} x_1\\
x_2\\
x_3
\end{array}}&
{ 
\left[\begin{array}{ccc}
2 &  1 &  2 \\
2 &  2 &  1 \\
1 &  2 &  2 
\end{array}\right].
}
\end{array}}
\end{array}
$$
The conditions in Theorem 1 are not fulfilled. However, none will be whipped by `Rank and Yank' since the final aggregation outcome is a cycle, i.e., $x_1 \succ x_2 \succ x_3\succ x_1$.

In the end of this section we note that, according to Black's stratege II (Black 1976) or Young (1974) one can see that, when condition (4) is satisfied, the alternatives have identical Borda counts.

\section{Concluding remarks}

In this paper we posed a question (i.e., conditions for none to be whipped by Rank and Yank) in social choice theory which is independent of two much discussed topics regarding conditions on social preference transitivity and on social choice set existence. We set forth some sufficient conditions. In a sense, the topic of `none to be whiped by Rank and Yank' is related to the stability of the organization. One can see that the condition of dual probability relation (Eq.(3)) specifies a certain balance in the probabilities of alternatives being ranked at positions. It appears that the organization's stability is somewhat related to the balance exhibited in the probabilities of alternatives occupying ranking positions.

Researchers have recognized for over 40 years that the majority cycles are relavant to the stability of political systems (see, e.g., Miller 1983, Ordeshook 1992, and many others). Therefore, the presented results have significance for the discussion of political stability in light of majority cycles. When the evaluation data are cardinal numbers, it is fairly simple to obtain the stability condition. But if the evaluation data are assumed of ordinal rankings, as discussed in this paper, things can be a bit more complicated. The presented conditions are merely sufficient for some particular specific cases. There are still many things that need further investigation. Future research might include, but not limited to, seeking more sufficient conditions, necessary conditions, necessary and sufficient conditions in certain circumstances, and their practical implicatioins to organization stability. In addition, the likelihood of the occurrence of `none to be whipped by Rank and Yank' under those assumptions made in this paper deserves further effort as well. We would like to leave them open to the readers.

\vspace{0.1cm}



\end{document}